\documentclass[twocolumn,preprintnumbers,amsmath,amssymb,superscriptaddress]{revtex4-1}
\usepackage{epsfig,bm}
\usepackage{ulem}
\usepackage{graphicx}
\usepackage{amsmath,color}
\begin{document}
\title{Correlated-Gaussian approach to  linear-chain states \\
-- Case of four $\alpha$-particles --}
\author{Y. Suzuki}
\affiliation{Department of Physics, Niigata University, Niigata 950-2181, Japan}
\affiliation{RIKEN Nishina Center, Wako 351-0198, Japan}
\author{W. Horiuchi}
\affiliation{Department of Physics, Hokkaido University, Sapporo 060-0810, Japan}
\date{\today}
\begin{abstract}
We show that correlated Gaussians with good 
angular momentum and parity provide flexible basis functions  
for specific elongated shape. 
As its application we study linear-chain states of four $\alpha$-particles 
in variation-after-projection calculations in which all the matrix 
elements are evaluated analytically. 
We find possible chain states for $J^{\pi}=0^+, 2^+, 4^+$ and perhaps $6^+$ 
with the bandhead energy being about 33 MeV from the ground state of $^{16}$O. 
No chain states with $J\geq 8$ are found. The nature of the rotational 
sequence of the chain states is clarified in contrast to a 
rigid-body rotation. The quadrupole deformation 
parameters estimated from the chain states increase from 0.59 to 1.07 for 
$2^+$ to $6^+$. This work suggests undeveloped fields for 
the correlated Gaussians 
beyond those problems which have hitherto been solved successfully. 
\end{abstract}
\maketitle

\section{Introduction}
\label{intro}

Spatially localized motion of nucleons  
plays an important role in nuclear structure through 
excitation mechanism, pairing, and $\alpha$-clustering, etc. 
A spatially localized single-particle (sp) orbit is needed and 
conveniently represented 
by a Gaussian wave packet (GWP)  
\begin{align}
\phi^{\nu}_{\bm s}(\bm r)=\left(\frac{\nu}{\pi}\right)^{\frac{3}{4}} 
e^{-\frac{\nu}{2}(\bm r-\bm s)^2},
\label{gwp}
\end{align} 
where $\bm s$ denotes the position of the packet and  $\nu^{-1/2}$ determines 
its width or spatial extension. Several acronyms used in this paper are 
listed in Appendix~\ref{glossary.acronym}. The GWP is 
widely used in the cluster model or its extended models~\cite{brink,fmd,amd}, 
but not popular in shell-model or configuration interaction 
calculations probably because it has no definite orbital angular 
momentum. As a localized orbit with good angular momentum, we propose 
a locally peaked Gaussian (LPG) specified by $k$ and $a$,  
\begin{align}
\varphi_{klm}^a(\bm r)=\frac{1}{G_{2k+l}} \Big(\frac{a^3}{\pi}\Big)^{\frac{1}{4}} 
(\sqrt{a}r)^{2k+l} e^{-\frac{a}{2}r^2}Y_{lm}(\hat{\bm r}),
\label{nodelessho}
\end{align}
where $\hat{\bm r}$ is the direction of $\bm r$. See 
Appendix~\ref{glossary.symbol} for $G_{2k+l}$. To make the text compact, 
we also put some other symbols and definitions there without mentioning.  
We discuss a relationship between the LPG and the GWP in the next section. 

It would be very interesting if the LPG could be extended to functions 
describing an $N$-particle system. Its possible candidate, correlated 
Gaussian (CG), was actually proposed more than 20 years ago by K. Varga and 
one of the present authors (Y.S.)~\cite{varga95,book} by 
extending the spherical CG~\cite{boys60,singer60} to that including 
a rotational motion of the system. The CG is concisely expressed as 
\begin{align}
f^{uA}_{KLM}(\bm \rho)={\cal N}^{uA}_{KL}|\widetilde{u}\bm \rho|^{2K+L}Y_{LM}(\widehat{\widetilde{u}\bm \rho})e^{-\frac{1}{2}\widetilde{\bm \rho}A\bm \rho},
\label{def.CG}
\end{align}
where the column vector $\bm \rho$  comprises 
$N-1$ relative coordinates, 
$({\bm \rho}_1, \ldots, {\bm \rho}_{N-1})$. 
The CG is characterized by a column vector $u=(u_i)$ of $(N-1)$-dimension 
and a symmetric, positive-definite $(N-1)\times (N-1)$ 
matrix $A=(A_{ij})$. The symbol $\widetilde{ \ }$ indicates 
a transpose of a column vector or a matrix, and therefore 
$\widetilde{u}\bm \rho$ (the so-called global vector) and $\widetilde{\bm \rho}A\bm \rho$ are short-hand notations for 
\begin{align}
\widetilde{u}\bm \rho=\sum_{i=1}^{N-1}u_i {\bm \rho}_i,\ \ \ 
\widetilde{\bm \rho}A\bm \rho=\sum_{i,j=1}^{N-1} A_{ij}{\bm \rho}_i\cdot {\bm \rho}_j.
\end{align}
$K$ in Eq.~(\ref{def.CG}) is a non-negative integer and ${\cal N}^{uA}_{KL}$ is 
a normalization constant (see also Ref.~\cite{suzuki98}).   
A formal resemblance of Eqs.~(\ref{nodelessho}) and (\ref{def.CG}) is apparent. 

The purpose of this paper is to demonstrate that a 
linear-chain (LC) state comprising the GWPs arranged in a row 
can in fact be very well represented by the CG  
with a suitable choice of $K$, $A$, and $u$, and furthermore to 
apply to the case of four $\alpha$-particles in order to examine whether the LC 
state can exist or not in $^{16}$O. 
Although it has been successful in a number of few-body problems~\cite{vargaprl98,suzuki00,nemura02,horiuchi12,horiuchi14a,mikami14,mitroy13}, the CG 
has focused most of its application on structure 
described well with small $K$ values, e.g., 0, 1, and 2. 
We will open up a new 
application of the CG by compactly describing a strongly deformed 
state rotating with high angular momentum.

The LC structure in nuclei was proposed as a candidate for a strongly 
deformed state that may play an 
important role for some excited states especially 
in light nuclei~\cite{morinaga56}. 
An experimental 
search was done in $^{16}$O~\cite{chevallier67} but no firm confirmation was
made yet. A theoretical analysis of the decay scheme 
of the LC state was first made in Ref.~\cite{suzuki72}. The 4$\alpha$ decay 
of some excited states in $^{16}$O 
has been studied experimentally~\cite{freer95,freer04,curtis16}. 
Recently the possibility of nuclear LC states in $^{16}$O as well as in other 
light nuclei has attracted renewed interest both theoretically and 
experimentally~\cite{suhara10,suhara11,ichikawa11,yao14,suhara14,zhao15,iwata15,royer15, freer14,fritsch16, yamaguchi17}.

In Sec.~\ref{localized.sp} we first begin with an anatomy of the GWP from 
the angular-momentum content, and show that the LPG can be a very convenient 
and flexible sp orbit representing a spatial localization. 
In Sec.~\ref{N-particle.lcc} we prove that the CG, an extension of the 
LPG to many-particle functions, is versatile enough to simulate a strongly 
deformed LC configuration of $N$ particles. We give a simple prescription 
for determining the CG parameters to fit the LC configuration as 
accurately as possible. 
An application of the present formulation is worked out in 
Sec.~\ref{four.alpha.LC} to examine possible LC states in $^{16}$O. 
The energy of the LC configuration of four $\alpha$-particles is studied 
by changing its size or length of the system as well as the total orbital angular momentum. 
Conclusions are drawn in Sec.~\ref{conclusion}.

\section{Spatially localized single-particle orbits}
\label{localized.sp}
\subsection{Angular-momentum expansion of Gaussian wave packet}
\label{amp.gwp}

The GWP~(\ref{gwp}) contains many partial-waves. Its orbital 
angular-momentum content is analyzed as 
\begin{align}
\phi^{\nu}_{\bm s}(\bm r)=\sum_{lm}\sqrt{4\pi}b_{l}(\eta)\phi_{slm}^{\nu}(\bm r)Y^*_{lm}(\hat{\bm s}),
\label{exp.shiftG}
\end{align}
where $\phi_{slm}^{\nu}(\bm r)$ is a normalized shifted-Gaussian (SG), 
\begin{align}
\phi_{slm}^{\nu}(\bm r)=\frac{2e^{-\eta}}{b_l(\eta)}\Big(\frac{\nu^3}{\pi}\Big)^{\frac{1}{4}}i_l(\nu sr)\,e^{-\frac{\nu}{2}r^2}Y_{lm}(\hat{\bm r}),
\label{radial.sgf}
\end{align}
expressed in terms of the modified spherical Bessel function of the 
first kind~\cite{MathFunc}, $i_l(x)=\sqrt{\frac{\pi}{2x}}I_{l+\frac{1}{2}}(x)$.
A dimensionless quantity $\eta$  
\begin{align}
\eta=\frac{1}{2}\nu s^2
\label{eta}
\end{align}
is a measure of the spatial localization of the packet.

\begin{figure}
\begin{center}
\epsfig{file=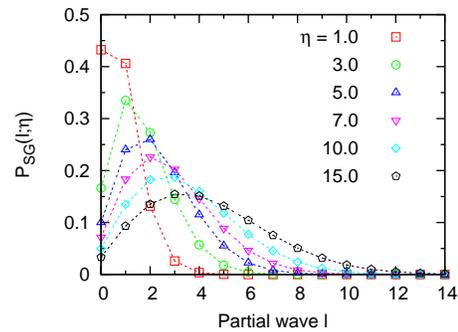,scale=1.0}
\caption{Probability of finding the component with partial wave $l$ in
the Gaussian wave packet characterized by $\eta$. See Eq.~(\ref{psg.single}). }
\label{gwp.am.exp}
\end{center}
\end{figure}

The probability of finding the component with angular momentum $l$ in the GWP  
is defined by 
\begin{align}
P_{\rm SG}(l;\eta)&=\sum_{m=-l}^l \frac{1}{4\pi} \int d\hat{\bm s} \, 4\pi [b_{l}(\eta)]^2 |Y_{lm}(\hat{\bm s})|^2\notag \\
&=(2l+1)[b_{l}(\eta)]^2,
\label{psg.single}
\end{align} 
which satisfies a sum rule, $\sum_{l=0}^{\infty} P_{\rm SG}(l;\eta)=1$. 
Figure~\ref{gwp.am.exp} plots $P_{\rm SG}(l;\eta)$ as a function of $l$ for some 
values of $\eta$. If $s$ is of the order of the nuclear surface $r_0A^{1/3}\,(r_0=1.1$\,fm, 
$A$ is the mass number), 
$\eta$ varies as $0.58A^{1/3}$ for the harmonic-oscillator (HO) choice of 
$\nu\approx 0.965A^{-1/3}$\,fm$^{-2}$. E.g.,  $\eta$ 
is about 1.5 for $A=16$ and 3.4 for $A=208$, respectively. On the other hand, 
if an $\alpha$-cluster described with its sp $\nu$ value 
(0.521\,fm$^{-2}$) is localized 
at $s=8$\,fm beyond the surface of $^{208}$Pb, $\eta$ increases 
to about 17, indicating the enhanced spatial localization. 
As shown in Fig.~\ref{gwp.am.exp}, the probability 
distribution extends to larger $l$ with increasing $\eta$, a consequence 
of the uncertainty relation between the angular momentum and 
the angular position. Many localized orbits with large $l$ are needed to represent 
the surface $\alpha$-clustering in $^{208}$Pb region~\cite{lovas98,varga92a,varga92b}.

Including spatially correlated configurations in the HO shell-model 
description is very tough because they require many major-shell 
excitations. Because of this 
even a large-scale shell-model calculation is not 
able to reproduce some cluster states in light nuclei~\cite{wloch05,maris09}. 
It is therefore important to develop simple sp 
orbits that are needed to construct such cluster states. 
Although a localized orbit like the SG could be a useful sp orbit, 
that is not a practically convenient 
basis function because calculations of various 
matrix elements are in general fairly involved. The LPG can instead be an 
ideal substitute as shown in the next subsection. 

\subsection{Locally peaked Gaussian}
\label{nodelessGauss}

We show that the LPG~(\ref{nodelessho}) well approximates the SG 
if $k$ and $a$ are appropriately chosen. 
The LPG has a merit that calculating matrix elements is easy. 
For example, the overlap between the LPG and the SG reads  
\begin{align}
\langle \phi^{\nu}_{slm}|\varphi^a_{klm}\rangle =O_l(\nu s, ak)
 e^{-\frac{a\eta}{\nu +a} } {_1}F_1(-k,\bar{l}+1;-\frac{\nu \eta}{\nu +a}),
\label{ovl.LPG.SG}
\end{align}
where $\bar{l}$ stands for $l+1/2$, and 
${_1}F_1$ is the confluent hypergeometric function~\cite{MathFunc}, which reduces 
to a 
polynomial for a non-negative integer $k$.  
To determine $k$ and $a$ that approximates a given SG as closely as possible, 
we require the expectation values of $r^2$ and $-\Delta$ calculated with the LPG, 
\begin{align}
&\langle \varphi^a_{klm}|r^2|\varphi^{a}_{klm} \rangle=
\left(2k+\bar{l}+1\right)\frac{1}{a},
\notag \\
&\langle \varphi^{a}_{klm} |-\Delta | \varphi^{a}_{klm} \rangle=
\Big(1+\frac{{\bar{l}}^{\,2}}{2k+\bar{l}}\Big)a,
\label{rkine.nodeless}
\end{align}
to be equal to the corresponding values of the SG,  
\begin{align}
&\langle \phi^{\nu}_{slm}|r^2|\phi^{\nu}_{slm}\rangle=
\Big(\bar{l}+1+\eta +\eta \frac{i_{l+1}(\eta)} {i_l(\eta)} \Big) \frac{1}{\nu}, \notag \\
&\langle \phi^{\nu}_{slm}|-\Delta|\phi^{\nu}_{slm}\rangle=\Big(\bar{l}+1-\eta+\eta \frac{i_{l+1}(\eta)}{i_l(\eta)}\Big){\nu}.
\label{me.r2.kine.SG}
\end{align}
This requirement is natural because the SG is characterized by its peak 
position and the falloff of the peak height. 
The condition leads to $k$ and $a$ as follows:
\begin{align}
k=\frac{1}{2}(z-\bar{l}), \ \ \ 
a=\frac{z+1}{\langle \phi^{\nu}_{slm}|r^2|\phi^{\nu}_{slm}\rangle},
\label{param.a.p}
\end{align}
where $z=\frac{1}{2}(b+\sqrt{D})$ with $b$ and $D$ being given by 
\begin{align}
&b=\Big(\bar{l}+1+\eta \frac{i_{l+1}(\eta)}{i_l(\eta)}\Big)^2-{\bar{l}}^{\,2}-1-\eta^2,\notag \\
&D=b^2-4\bar{l}^{\,2}.
\end{align}

\begin{figure}
\begin{center}
\epsfig{file=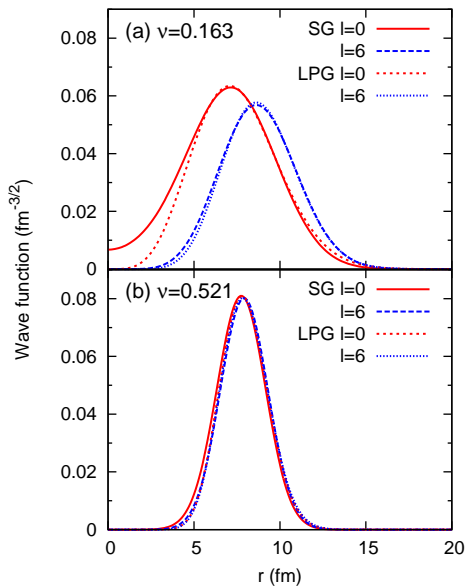,scale=1.0}
\caption{Comparison of the radial functions with $l=0$ and 6 between 
the SG, $\phi^{\nu}_{sl}(r)$, and the LPG, $\varphi^a_{kl}(r)$.   
The value of $s$ is set to 8\,fm. Both $\nu$ and $a$ are given in fm$^{-2}$.  
Case (a) $\nu=0.163$: $(a,k)=(0.0796, 2)$ for $l=0$ and $(0.1076, 1)$ 
for $l=6$. Case (b) $\nu=0.521$: 
$(a,k)=(0.2630, 8)$ for $l=0$ and $(0.2605, 5)$ for $l=6$. 
The overlap integral between the LPG and the SG is larger than 0.998 
in all the cases. }
\label{comp.sg.lpg}
\end{center}
\end{figure}

We have numerically checked that  
$b-2\bar{l}$ is non-negative, which guarantees 
that $k$ is non-negative. See Appendix~\ref{func.fzl} for this. 
For a practical purpose, $k$ is restricted to a non-negative integer 
closest to $(z-\bar{l})/2$.  
Once $k$ is fixed, $a$ is set to maximize the 
overlap~(\ref{ovl.LPG.SG}) between the arithmetic and geometric 
means of two $a$ values 
that reproduce the respective expectation values of $r^2$ and $-\Delta$ 
of the SG. 
Figure~\ref{comp.sg.lpg} compares $l=0$ and 6 radial functions between the LPG  
and the SG with $s=8$\,fm. In case (a), 
$\nu=0.163$\,fm$^{-2}$ is the HO size parameter appropriate in $^{208}$Pb 
region, whereas $\nu=0.521$\,fm$^{-2}$ in case (b) reproduces the size of the 
$\alpha$-particle with the $(0s)^4$ configuration. $\eta$ is quite 
different; 5.2 in case (a) and 16.7 in case (b). In both cases the LPG 
very well approximates the SG.

\section{Linear-chain configurations of $N$ particles}
\label{N-particle.lcc}
\subsection{Gaussian wave-packet representation}
\label{gwprepre}

For the sake of simplicity, we assume that $N$ particles have an equal mass 
$m$. In the LC state they are 
conveniently described by the GWPs that are all 
centered at some positions on a straight line passing through the coordinate 
origin. This intrinsic state rotates around the center-of-mass (c.m.) 
of the system with high angular velocity. Let a unit vector 
$\bm e$ ($|\bm e|=1$) denote the direction of the line. 
The LC state with the total orbital angular momentum $L$ and its projection $M$ is defined by 
\begin{align}
\Phi_{LM}={\cal N}_L\int d\bm e\, Y_{LM}(\hat{\bm e})\prod_{i=1}^N \phi_{S_i{\bm e}}^{\nu}({\bm r}_i).
\label{LCS}
\end{align}
The $i$-th particle is centered at the position $S_i\bm e$ on the 
line. We assume that $\sum_{i=1}^N S_i=0$ to ensure that the c.m. 
motion of the system remains fixed around the origin. ${\cal N}_L$ is a 
normalization constant given by 
where $H$ is an analogue of the localization measure $\eta$~(\ref{eta}): 
\begin{align}
H=\frac{1}{2}\nu \sum_{i=1}^NS_i^2=\frac{1}{2}\nu \widetilde{S}S.
\end{align}
Here $S=(S_i)$ is an $N$-dimensional column vector.

The c.m. motion of $\Phi_{LM}$ is separated by transforming the sp 
coordinates $\bm r=({\bm r}_i)$ to a set of the 
relative coordinates, $\bm \rho=({\bm \rho}_i)$, and the c.m. coordinate 
${\bm R}_{N}$:
\begin{align}
\left(
\begin{array}{c}
\bm \rho \\
{\bm R}_N\\
\end{array}
\right)
= U\bm r,\ \ \  \bm r=U^{-1}
\left(
\begin{array}{c}
\bm \rho \\
{\bm R}_N\\
\end{array}
\right).
\label{coord.trans}
\end{align}
If $\bm \rho$ is a set of Jacobi coordinates, $U$ and $U^{-1}$ are 
\begin{align}
&U=\left(
\begin{array}{ccccc}
-1 & 1 & 0 & \cdots & 0 \\
-\frac{1}{2} & -\frac{1}{2} & 1 & \cdots & 0 \\
\vdots & & & & \vdots \\
-\frac{1}{N-1} & -\frac{1}{N-1} & \cdots & \cdots & 1 \\
\frac{1}{N} & \frac{1}{N} & \cdots & \cdots & \frac{1}{N} \\
\end{array}
\right),\notag \\
\notag \\
&U^{-1}=\left(
\begin{array}{ccccc}
-\frac{1}{2} & -\frac{1}{3} & \cdots & -\frac{1}{N} & 1 \\
\frac{1}{2} & -\frac{1}{3} & \cdots & -\frac{1}{N} & 1 \\
\vdots & & & & \vdots \\
0 & 0 & \cdots & \cdots & 1 \\
0 & 0 & \cdots & \frac{N-1}{N} & 1 \\
\end{array}
\right).
\end{align} 
Let $U_J$ denote the first $(N-1)\times N$ submatrix of 
$U$ and $U_J^{\ -1}$ the first $N\times (N-1)$ submatrix of $U^{-1}$. 
Note the following identities  
\begin{align}
&U_J\widetilde{U_J}=\Lambda^{-1},  \ \ \ \widetilde{U_J^{\ -1}}U_J^{\ -1}=\Lambda, \notag \\
&U_J^{-1}\Lambda^{-1} \widetilde{U_J^{-1}}=U_J^{-1}U_J=1_N-\frac{1}{N}E_N,
\end{align}
where $1_N$ is the $N\times N$ unit matrix, $E_N$ is the $N\times N$ matrix unit whose elements are 
all unity, and $\Lambda$ is an $(N-1)\times (N-1)$ diagonal matrix defined by 
\begin{align}
\Lambda_{ij}=\frac{i}{i+1}\delta_{i,j}.
\end{align}
The LC state~(\ref{LCS}) reduces to a product of the c.m. part and the 
normalized intrinsic part $\Phi^{\nu}_{SLM}({\rm LC})$:
\begin{align}
\Phi_{LM}
=\phi_{\bm 0}^{N\nu}({\bm R}_N)\Phi^{\nu}_{SLM}({\rm LC}),
\end{align}
where 
\begin{align}
&\Phi^{\nu}_{SLM}({\rm LC})\notag \\
&=\frac{2e^{-H}}{b_L(H)}\Big(\frac{\nu^{3N-3}}{N^3\pi^{3N-5}}\Big)^{\frac{1}{4}}
e^{-\frac{1}{2}\widetilde{\bm \rho}A_0\bm \rho}
i_L(|\widetilde{u_0}\bm \rho|)Y_{LM}(\widehat{\widetilde{u_0}\bm \rho})
\label{chain.wf}
\end{align}
with
\begin{align}
A_0= \nu \Lambda,\ \ \  u_0=\nu \widetilde{U_J^{-1}}S.
\label{Au.param}
\end{align}
Note that the parity of $\Phi^{\nu}_{SLM}({\rm LC})$ is $(-1)^L$. 

Some basic operators are conveniently expressed in terms of $\bm \rho$. 
For example, 
\begin{align}
&\bm r_i-{\bm R}_N=\sum_{k=1}^{N-1} (U_J^{-1})_{ik}{\bm \rho}_k,\\
&{\bm r}_i-{\bm r}_j=\sum_{k=1}^{N-1} \omega^{(ij)}_k {\bm \rho}_k=\widetilde{\omega^{(ij)}}\bm \rho
\label{rel.distance}
\end{align}
with
\begin{align}
 \omega^{(ij)}_k=(U^{-1}_J)_{ik}-(U^{-1}_J)_{jk}.
\end{align}
The hyperradius $R$ of the system is defined by
\begin{align}
R^2=\sum_{i=1}^{N}({\bm r}_i-{\bm R}_N)^2=\sum_{i=1}^{N-1}\frac{i}{i+1}{\bm \rho}_i^2=\widetilde{\bm \rho}\Lambda{\bm \rho}.
\end{align}
The kinetic energy with the c.m. kinetic-energy $T_{\rm c.m.}$ 
being subtracted reads 
\begin{align}
T_{\rm in}=\sum_{i=1}^N \frac{{\bm p_i^2}}{2m}-T_{\rm c.m.}=\sum_{i=1}^{N-1}\frac{i+1}{2mi}{\bm \pi}_i^2=\frac{1}{2m}\widetilde{\bm \pi}\Lambda^{-1} {\bm \pi}, 
\end{align}
where ${\bm \pi}=({\bm \pi}_j)$ with 
${\bm \pi}_j=-i\hbar {\partial}/{\partial \bm \rho}_j$ is a column vector 
of $(N-1)$-dimension.

The LC state~(\ref{chain.wf}) takes a form quite similar to 
the SG~(\ref{radial.sgf}). The argument of $i_l(\nu sr)$ becomes $\nu s^2=2\eta$ 
at the peak of the SG. Likewise,  
since $\Phi_{LM}$ is peaked at $\bm r=S\bm e$, $\Phi^{\nu}_{SLM}({\rm LC})$ is 
peaked at $\bm \rho=U_JS\bm e$. For this $\bm \rho$, the argument of 
$i_L(\widetilde{u_0}\bm \rho|)$ becomes 
$|\widetilde{u_0}\bm \rho|=\nu \widetilde{S}S|\bm e|=2H$. 
In parallel to the sp SG case, $H$ is indeed the 
localization measure for the $N$-body LC state. This analogy becomes 
furthermore 
substantial by calculating the expectation values of $R^2$ and $T_{\rm in}$ 
(cf. Eq.~(\ref{me.r2.kine.SG})):
\begin{align}
&\langle \Phi^{\nu}_{SLM}({\rm LC})|R^2| \Phi^{\nu}_{SLM}({\rm LC}) \rangle \notag \\
&=\frac{1}{\nu}\Big[\frac{3}{2}(N-1)+L+H +H\frac{i_{L+1}(H)}{i_L(H)}\Big],\notag \\
&\langle \Phi^{\nu}_{SLM}({\rm LC})| T_{\rm in} | \Phi^{\nu}_{SLM}({\rm LC}) \rangle \notag \\
&=\frac{\hbar^2\nu}{2m}\Big[\frac{3}{2}(N-1)+L-H+H\frac{i_{L+1}(H)}{i_L(H)}\Big].
\label{radius.kinetic.LC}
\end{align}
This kinetic-energy expectation value  
gives approximate $L(L+1)$ dependence up to large $L$ already 
for $H \geq 30$ in spite of 
its opacity. See Appendix~\ref{func.fzl} for some detail. 

\subsection{Correlated-Gaussian approximation}
\label{GWP-CG.appr.}

We have shown that in one-variable case the LPG~(\ref{nodelessho}) 
very well approximates 
the SG~(\ref{radial.sgf}) and also that the functional form of the 
$N$-particle LC state~(\ref{chain.wf}) is similar to that of the SG. 
Here we show that the LC state can be very well 
approximated by the CG~(\ref{def.CG}), which is a natural extension of 
the LPG to the many-variable case.

Various matrix elements with the CG can easily be obtained by making 
use of its generating function $g$~\cite{varga95}:
\begin{align}
f^{uA}_{KLM}(\bm \rho)&=\frac{{\cal N}^{uA}_{KL}}{B_{KL}}\int d{\bm e}Y_{LM}(\hat{\bm e}) \notag \\
&\times \Big(\frac{d^{2K+L}}{d\alpha^{2K+L}}g(\alpha,\bm e;u,A,\bm \rho)\Big)_{\alpha=0},
\end{align}
where $\bm e$ is a unit vector. 
The overlap of Eqs.~(\ref{chain.wf}) and (\ref{def.CG}) 
is (see Ref.~\cite{suzuki98} for detail)
\begin{align}
&\langle f^{uA}_{KLM}|\Phi^{\nu}_{SLM}({\rm LC})\rangle\notag \\
&=\frac{e^{-H}}{b_L(H)}\Big(\frac{\det 4\nu A}{N(\det B)^2}\Big)^{\frac{3}{4}}
 \frac{e^{p_0}}{\sqrt{{\bar p}^{2K+L}}} G_{KL}(p,q),
\label{ovl}
\end{align}
where the matrix $B$ is $B=A+A_0$ and 
\begin{align}
&\bar p=\frac{1}{4}\widetilde{u}A^{-1}u,\ \ \ 
p=\frac{1}{2}\widetilde{u}B^{-1}u,\notag \\
&q=\widetilde{u}B^{-1}u_0,\ \ \ p_0=\frac{1}{2}\widetilde{u_0}B^{-1}u_0.
\end{align}

To determine the CG parameters, ($A$, $u$, $K$), that 
well approximate $\Phi^{\nu}_{SLM}({\rm LC})$ characterized by ($\nu$, $S$) or 
($A_0, u_0$), we follow the same route as that in Sec.~\ref{nodelessGauss}. 
The expectation values~(\ref{radius.kinetic.LC}) are compared to 
those with the CG~(\ref{def.CG})~\cite{suzuki98}: 
\begin{align}
&\langle f^{uA}_{KLM}|R^2
|f^{uA}_{KLM}\rangle 
=\frac{3}{2}{\rm Tr} A^{-1}\Lambda + (L+2K)\frac{\bar{q}}{\bar{p}},\notag \\
&\langle f^{uA}_{KLM}|T_{\rm in}|f^{uA}_{KLM}\rangle \notag \\
&=\frac{\hbar^2}{2m}\Big[ \frac{3}{2}{\rm Tr}A\Lambda^{-1}+(L-2K+4C_{KL})\frac{\bar{\lambda}}{\bar{p}}\Big],
\label{rms.kinetic.cg}
\end{align}
where  
\begin{align}
&\bar{q}=\frac{1}{4}\widetilde{u}A^{-1}\Lambda A^{-1}u, \ \ \ 
\bar{\lambda}=\frac{1}{4}\widetilde{u}\Lambda^{-1} u,\notag \\
&C_{KL}=\frac{1}{\gamma_{KKL}(1)} \gamma_{KKL}^{\prime}(1).
\end{align}
Here $\gamma_{KK'L}^{\prime}(x)=\frac{d}{dx}\gamma_{KK'L}(x)$. 
As a simplest choice, let us assume that $A$ is proportional to $A_0$: 
\begin{align}
A=aA_0={a}\nu \Lambda. 
\label{A.choice}
\end{align}
The condition to be satisfied then reads 
\begin{align}
&\frac{1}{a}\left[\frac{3}{2}(N-1)+L+2K\right]\notag \\
& \qquad =\frac{3}{2}(N-1)+L+H+H\frac{l_{L+1}(H)}{i_L(H)},\notag \\
&a\left[\frac{3}{2}(N-1)+L-2K+4C_{KL}\right]\notag \\
& \qquad =\frac{3}{2}(N-1)+L-H+H\frac{i_{L+1}(H)}{i_L(H)}.
\label{comp.r2t}
\end{align}
$K$ is determined by requiring the product of the left-hand sides of 
Eq.~(\ref{comp.r2t}), which is a function of $K$ and independent of $a$, 
to be equal to that of the right-hand sides. 
Since $K$ is set to a non-negative integer, the condition may not be  
perfectly met but $K$ is fixed so as to satisfy the condition as much
as possible. For this $K$ we have two $a$ values, one determined from 
the first equation in Eq.~(\ref{comp.r2t}) and the other determined from the 
second equation. 
Both values are found to be almost equal and we choose $a$ as 
an arithmetic average of those two values.   
Note that $K$ and $a$ or $A$ are determined depending on $L, H$, and $N$ but 
independent of $u$. 

Once $A$ and $K$ are set, the overlap~(\ref{ovl}) depends on $u$  
only through the term,  
\begin{align}
\frac{1}{\sqrt{\bar{p}^{2K+L}}}G_{KL}(p,q)=\Big(\frac{2a}{a+1}\Big)^{\frac{1}{2}(2K+L)}G_{KL}(1,z)
\label{max.ovl}
\end{align}
with 
\begin{align}
z=\sqrt{\frac{2}{(a+1)\nu}}\frac{\widetilde{u}\Lambda^{-1} u_0}{\sqrt{ \widetilde{u}\Lambda^{-1} u}}. 
\end{align}
The overlap becomes a maximum when $G_{KL}(1,z)$ or $z$ reaches a maximum. The maximum of $z$ occurs for such $u$ that is proportional to  $u_0$, {\it i.e.}, 
Max $(z)=\sqrt{4H/(a+1)}$. For definiteness, $u$ is set equal to $u_0$. 
In this way the CG that has the maximum overlap 
with $\Phi^{\nu}_{SLM}({\rm LC})$ is determined to be 
$f^{u_0\, aA_0}_{KLM}(\bm \rho)$, which is denoted 
$f^{\nu}_{SLM}({\rm LC:CG})$ in order to emphasize its LC character. 
As shown in Eq.~(\ref{Au.param}), $A_0$ is unique but $u_0$ depends on 
the column vector $S$ for a given $H$. The CGs with different $u_0$ 
parameters all have the same maximum overlap with 
$\Phi^{\nu}_{SLM}({\rm LC})$. 
Table~\ref{CGGV.LCS} lists the CG parameters, $K$ and $a$, determined 
in this way together with the maximum overlap,  
$\langle \Phi^{\nu}_{SLM}({\rm LC})|f^{\nu}_{SLM}({\rm LC:CG}) \rangle$ 
for some sets of $H$ and $L$ values. 
Observing that it is close to unity, we conclude that the LC 
configuration can be well approximated with the CG~(\ref{def.CG}) 
provided its parameters are determined as mentioned above. 
It is worth while stressing that the CG approximation works excellently 
even for extremely large $L$. Numerical angular momentum projection for 
such $L$ states may be tough in general. No such difficulty arises here  
thanks to the analytic manipulation.

\begin{table*}
\caption{Maximum overlap, $\langle \Phi^{\nu}_{SLM}({\rm LC})|f^{\nu}_{SLM}({\rm LC:CG}) \rangle$, between the CG  and the LC state with the localization measure $H$ for a system of four particles ($N=4$). $K$ and $a$ (or $A$) are the CG parameters that  
maximize the overlap with the LC state.  }
\label{CGGV.LCS}
\begin{center}
\begin{tabular}{lccccccccccccccc}
\hline\hline
$L \diagdown H$&& 10 &  20 & 30 & 40 & 50 & 60 & 70 & 80 & 90 & 100 & 140 & 180 & 220 & 260 \\
\hline 
0 &$K$ &  8 & 17 & 25 & 34 & 43 & 52 & 60 & 69 & 78 & 87 & 122 & 157 & 192 & 227 \\
& $a$ &  0.872 & 0.879 & 0.866 & 0.871 & 0.874 & 0.877 & 0.871 & 0.873 & 0.875 & 0.876 & 0.876 & 0.876 & 0.875 & 0.875\\
&Overlap &  0.973 & 0.974 & 0.973 & 0.974 & 0.974 & 0.974 & 0.974 & 0.974 & 0.974 & 0.974 & 0.974 & 0.974 & 0.974 & 0.974 \\
\hline
10 &$K$ &  6 & 13 & 22 & 30 & 39 & 47 & 56 & 65 & 73 & 82 & 117 & 152 & 187 & 222\\
& $a$ &  0.933 & 0.880 & 0.891 & 0.878 & 0.882 & 0.874 & 0.876 & 0.878 & 0.873 & 0.875 & 0.875 & 0.875 & 0.875 & 0.875\\
&Overlap &  0.986 & 0.978 & 0.976 & 0.975 & 0.975 & 0.975 & 0.975 & 0.975 & 0.974 & 0.974 & 0.974 & 0.974 & 0.974 & 0.974\\
\hline
20 &$K$ &  5 & 12 & 19 & 27 & 35 & 44 & 52 & 61 & 69 & 78 & 113 & 147 & 182 & 217\\
& $a$ &  0.940 & 0.924 & 0.893 & 0.887 & 0.879 & 0.885 & 0.878 & 0.881 & 0.875 & 0.878 & 0.878 & 0.874 & 0.875 & 0.875\\
&Overlap &  0.993 & 0.985 & 0.981 & 0.978 & 0.977 & 0.976 & 0.976 & 0.975&0.975 & 0.975 & 0.975 & 0.974 & 0.974 & 0.974\\
\hline
30 &$K$ &  5 & 11 & 18 & 25 & 33 & 41 & 49 & 57 & 66 & 74 & 109 & 143 & 178 & 213\\
& $a$ &  0.966 & 0.936 & 0.920 & 0.899 & 0.895 & 0.890 & 0.884 & 0.879 & 0.883 & 0.878 & 0.880 & 0.876 & 0.876 & 0.876\\
&Overlap &  0.996 & 0.990 & 0.985 & 0.982 & 0.980 & 0.978 & 0.977 & 0.977 &0.976 & 0.976 & 0.975 & 0.975 & 0.975 & 0.975\\
\hline\hline
\end{tabular}
\end{center}
\end{table*}

The overlap of two CGs has simple dependence on their parameters: 
\begin{align}
&\langle f^{uA}_{KLM}|f^{vB}_{K'LM}\rangle 
= \frac{\gamma_{KK'L}(\bar{t})\,{\bar{t}}^{\frac{L}{2}}} {\sqrt{\gamma_{KKL}(1)\gamma_{K'K'L}(1)}}\notag \\
&\quad \times 
\Big(\frac{\sqrt{\det AB}}{\det C}\Big)^{\frac{3}{2}}\Big(\frac{\widetilde{u}C^{-1}u}{\widetilde{u}A^{-1}u}\Big)^K
\Big(\frac{\widetilde{v}C^{-1}v}{\widetilde{v}B^{-1}v}\Big)^{K'},
\label{ovlap.CG}
\end{align}
where 
\begin{align}
C=\frac{1}{2}(A+B),\ \ \ \ \ {\bar{t}}=\frac{(\widetilde{u}C^{-1}v)^2}
{(\widetilde{u}A^{-1}u)\, (\widetilde{v}B^{-1}v)}.
\end{align}
As shown in the table, $f^{\nu}_{SLM}({\rm LC:CG})$ may have very large $K$. 
On the other hand, low-lying states are described well with the CGs 
with small or even $K=0$ values. 
A specific overlap, $\langle f^{uA}_{KLM}|f^{uA}_{0LM}\rangle$, reduces to 
$1/\sqrt{\gamma_{KKL}(1)}$, and becomes very small 
for very large $K$.

\subsection{Geometrical shape and deformation}
\label{shape}

Angular-momentum projection is often carried out after a variational 
calculation is first performed by using unprojected basis functions.  
This is the so-called variation before projection (VBP). 
It is of course desirable to perform the projection 
before variation, that is, the variation after projection (VAP). 
Since the angular-momentum projection usually takes expensive 
computer-time in the numerical integration over the Euler angles 
(see, e.g., Ref.~\cite{enyo98}),
the VBP is employed in most calculations. 
Or even the full angular-momentum projection
is not made but is treated 
in a cranking model approximation~\cite{flocard84,ring80,nilsson95}. 
The CG already carries good angular momentum, 
making the VAP calculation very easy. 

In the VBP calculation, the geometrical shape of the unprojected configuration 
is often discussed. Although such shape is not ``observable'', the VBP basis 
functions give an intuitive image of the 
state, as shown in, e.g., 
Refs.~\cite{suhara10,suhara11,ichikawa11,yao14,suhara14,zhao15}. 
An elegant way to extract the intrinsic density from the VAP 
wave function was shown in Ref.~\cite{wiringa00}. Here we discuss a simpler way 
to get the geometrical picture from the CG.  
To characterize the intrinsic shape, we can make use of a set of operators, 
$r_{ij}^2$ and $(\bm r_j-\bm r_i)\cdot(\bm r_k-\bm r_j)$ $(i<j<k)$, for all 
pairs, and let $D_{ij}^2$ and $D_{ij}D_{jk}\cos \Theta_{ijk}$ denote 
their expectation values. All sets of $D_{ij}$ and 
$\Theta_{ijk}$ serve to extract the shape. Note that those 
operators are all expressed in a concise 
form, $\widetilde{\bm \rho}\Omega {\bm \rho}$, with the 
$(N-1)\times (N-1)$ matrix $\Omega$ being $\omega^{(ij)}\widetilde{\omega^{(ij)}}$ and $\omega^{(ji)}\widetilde{\omega^{(kj)}}$, respectively. 
See Eq.~(\ref{rel.distance}). The matrix element of  
$\widetilde{\bm \rho}\Omega {\bm \rho}$ is obtained in exactly the same way as 
that of $R^2$. We give it in Appendix~\ref{me.rhoOmegarho} for convenience. 

To get information on the deformation of the VAP wave function $\Psi_{LM}$, 
we may use the matrix element of the (mass) quadrupole moment $Q$, 
\begin{align}
\langle \Psi_{LL}|Q|\Psi_{LL} \rangle,
\end{align}
where
\begin{align}
Q=3R_z^2 - R^2.
\end{align}
Here $R_z^2=\sum_{i=1}^N(z_i-Z_N)^2$ with $z_i-Z_N$ being the $z$ component 
of ${\bm r}_i-{\bm R}_N$. We define a quantity $\delta_2$
\begin{align}
\delta_2=\frac{\langle \Psi_{LL} |Q|\Psi_{LL}\rangle}{ \langle \Psi_{LM}|R^2 | \Psi_{LM}\rangle},
\label{delta2}
\end{align}
which leads to the following relation
\begin{align}
\frac{\langle x^2 \rangle + \langle y^2 \rangle }{\langle z^2 \rangle}=\frac{2-\delta_2}{1+\delta_2},
\end{align}
where, e.g., $\langle z^2 \rangle$ stands for $\langle \Psi_{LL} |R_z^2|\Psi_{LL}\rangle$, and we discuss the deformation of the LC state 
in Sec.~\ref{search.LC}. Appendix~\ref{me.quadrupole} 
gives a formula to calculate the matrix element 
$\langle f^{uA}_{KLL}|Q|f^{vB}_{K'LL} \rangle$.

\section{Four-$\alpha$ linear-chain states}
\label{four.alpha.LC}

\subsection{Potential parameters}

A two-body $\alpha$-$\alpha$ potential we use here 
is the same as used in Ref.~\cite{suno16}.  
It consists of nuclear ($V_{\rm 2B}$) and Coulomb  ($V_{\rm C}$) terms:
\begin{align}
v_{ij}&=125 \exp\Big(-\frac{r_{ij}^2}{1.53^2}\Big)-30.18\exp\Big(-\frac{r_{ij}^2}{2.85^2}\Big)\notag \\
&+\frac{4e^2}{r_{ij}} {\rm erf}(0.60141 r_{ij}).
\end{align}
Energy and length are given in units of MeV and fm, respectively. 
A three-$\alpha$ nuclear potential, $V_{\rm 3B}=\sum_{i<j<k}v_{ijk}$, is also 
introduced. We may express it as 
\begin{align}
v_{ijk}=v(ij)v(jk)v(ki)
\end{align}
with
\begin{align}
v(ij)=v_r\exp\Big(-\frac{r_{ij}^2}{\rho_r^2}\Big)+v_a\exp\Big(-\frac{r_{ij}^2}{\rho_a^2}\Big).
\end{align}
$v(ij)$ is assumed to be $J^{\pi}$-independent. 
The parameters used in Ref.~\cite{suno16} are  
$v_r=0$, $v_a=-5.49$\,MeV$^{1/3}$, $\rho_a=3.395$, if $v_a$ is replaced by 
the average of the $0^+$ and $2^+$ strengths. With the mass of the 
$\alpha$-particle and the charge constant, $\hbar^2/m=10.5254$\,MeV\,fm$^2$, 
$e^2=1.43996$\,MeV\,fm, the energies of the 3$\alpha$ $0^+$ 
ground state and the first excited $2^+$ state calculated with that  
three-body potential are 
respectively about $-10.9$ and $-1.6$\,MeV, which are 
compared to the experimental 
values of $^{12}$C, $-7.28$ and $-2.84$\,MeV. These energies are obtained 
by using the CGs~(\ref{def.CG}), in which the $2\times 2$  
matrix $A$ is provided with the ansatz  
\begin{align}
\widetilde{\bm \rho}A{\bm \rho}=\sum_{j>i=1}^3 \frac{(\bm r_i-\bm r_j)^2}{b_{ij}^2},
\end{align}
where $b_{ij}$ is chosen for each $i,j$ in a geometric progression 
as $b_0 p^{n-1}$ ($n=1,\ldots,N_p$). Also non-zero $K$ values with $K \leq 2$ 
are allowed. The energy depends on the parameters $b_0, \, p$ and 
$N_p$ as well as $u$. A slight improvement is possible 
by including the short-ranged repulsive force with nonzero $v_r$, the existence of which is 
physically reasonable considering the Pauli principle acting between the 
$\alpha$-$\alpha$ relative motion~\cite{saito77}. 
With the parameters,
$v_r=6.5$\,MeV$^{1/3}$, $\rho_r=1.43$, $v_a=-6.0$\,MeV$^{1/3}$, 
$\rho_a=3.40$,
the $0^+$ and $2^+$ energies turn out to be about 
$-8.7$ and $-1.4$\,MeV. In what follows we use this three-body potential. 

The width parameter $\nu$ of the GWP specifies the spatial 
extension of the c.m. motion of the $\alpha$-particle. It appears in 
the LC configuration~(\ref{chain.wf}), or its approximated CG. 
Since it is four times the sp HO parameter of 
the $\alpha$-particle, we set $\nu$ to  
2.084\,fm$^{-2}$. The energy we calculate is to be taken 
the one measured from the $4\alpha$ threshold ($E_{\rm th}=14.436$\,MeV from the ground 
state of $^{16}$O).

\subsection{Features of correlated-Gaussian calculations}

Incorporating the boson symmetry of $\alpha$-particles in the CG formalism is 
very easy~\cite{varga95,book}. 
The permutation $P=\begin{pmatrix} 1& 2 & \ldots & N \\ P_1 & P_2 & \ldots &
P_{N} \end{pmatrix}$ changes $\bm r_i \to \bm r_{P_i}$ $(i=1,\ldots, N)$,
namely  $\bm r \to {\cal P}\bm r$, with the $N\times N$ matrix ${\cal P}$ being defined by ${\cal P}_{ij}=\delta_{j,P_i}$. With this permutation, 
$\bm \rho=U_J\bm r$ undergoes the transformation, $\bm \rho \to U_J{\cal P}\bm r$. 
Substitution of $\bm r=U_J^{-1}\bm \rho$ proves that 
$P$ transforms $\bm \rho$ to 
$U_J{\cal P}U_J^{-1}\bm \rho \equiv T_P\bm \rho$. 
Thus the CG~(\ref{def.CG}) is subject to the following change:  
$Pf^{uA}_{KLM}(\bm \rho)=f^{\widetilde{T_P}u\, \widetilde{T_P}A{T_P}}_{KLM}(\bm \rho)$,  
that is, the permutation $P$ sets the CG to a CG with 
$A$ and $u$ being replaced by $\widetilde{T_P}A{T_P}$ and $\widetilde{T_P}u$. 
In exactly the same way, a different choice of the relative coordinate 
set can be very easily incorporated in the CG formalism.

We evaluate the CG matrix elements using the formula given 
in Ref.~\cite{suzuki98}. To calculate the matrix element of the 
Gaussian-type potential
we can use a much simpler route as follows. The use of 
Eq.~(\ref{rel.distance}) enables us to express $v_{ij}=e^{-a r_{ij}^2}$ as 
$e^{-a {\widetilde{\bm \rho}} \Omega^{(ij)}{\bm \rho}}$ with 
$\Omega^{(ij)}=\omega^{(ij)} \widetilde{\omega^{(ij)}}$. 
The matrix element of the 
potential thus reduces to that of overlap type. 
The three-body force of Gaussian radial form is also 
treated in exactly the same way as the two-body case.  
The matrix element of the Coulomb potential is calculated by 
applying the above result. The Coulomb potential,  
$v_C(r)={\rm erf}(\beta r)/r$, is expressed as an integral of 
the Gaussian-type potential
\begin{align}
v_C(r)=\frac{2\beta}{\sqrt{\pi}} \int_0^1 \, dz\, e^{-\beta^2z^2 r^2}, 
\end{align}
and we reduce its matrix element to that 
of the Gaussian-type potential with a variable range parameter $a=\beta^2 z^2$.

\subsection{Arrangements of four $\alpha$-particles}
\label{arrangements}

Searching for 4$\alpha$ LC states requires a careful study 
on their energies with respect to the angular 
momentum and chain length. We examine the energy of the LC configuration 
with $J^{\pi}=L^+$ (even $L$) by changing $H$ or equivalently
root-mean-square (rms) radius. It is important to 
get a global change of the system's energy with respect to that key parameter~\cite{nielsen01,suzuki15}.

For a given $H$ there are different sets of $S$ denoted 
$S_{\kappa}^H$. Each $S_{\kappa}^H$ defines the LC configuration that is 
very well approximated by the CG as shown in Sec.~\ref{GWP-CG.appr.}. 
The set $S_{\kappa}^H$ corresponds to the vibration of $\alpha$-particles  
along the line of the LC state. Possible independent sets are 
prepared as follows. By eliminating one of the 
elements of $S$, say, $S_4=-(S_1+S_2+S_3)$, $H$ reads 
\begin{align}
H&=\nu(S_1^2+S_2^2+S_3^2+S_1S_2+S_1S_3+S_2S_3)\notag \\
&=\nu \widetilde{\varsigma}{\cal M}{\varsigma},
\end{align}
where $\varsigma$ is a column vector comprising the elements 
$S_1,S_2,S_3$ and ${\cal M}$ is a 3$\times$3 matrix with ${\cal M}_{ii}=1$, ${\cal M}_{i \neq j}=1/2$. The eigenvalues of ${\cal M}$ are 2, 1/2, and 1/2. With  
a suitable 3$\times$3 orthogonal matrix ${\cal T}$, $\varsigma={\cal T}Z$, 
$H$ can be recast to a quadratic form
\begin{align}
H=\nu \Big(2Z_1^2+\frac{1}{2}Z_2^2+\frac{1}{2}Z_3^2\Big).
\end{align}
By parametrizing $Z$ in terms of two angles, $\theta$ ($0\leqq \theta < 2\pi $) 
and $\phi$ ($0\leqq \phi < 2\pi $), as 
$Z_1=\sqrt{H/2\nu}\cos \theta$,  
$Z_2=\sqrt{2H/\nu}\sin \theta \cos \phi$, and  
$Z_3=\sqrt{2H/\nu}\sin \theta \sin \phi$, we can cover 
all possible vectors $S$. We discretize $\theta$ and $\phi$ 
in 5$^\circ$ mesh to generate $S_{\kappa}^H$, and allow those 
configurations  
that have mutual overlaps of  
less than 0.85 with others  in order to avoid possible linear dependence of 
the basis functions. 

The total wave function for the LC state is in general given as a superposition 
of different LC configurations of four $\alpha$-particles
\begin{align}
\Psi_{LM}=\sum_{H {\kappa}} C_{S^H_{\kappa}} {\cal S} f^{\nu}_{S^H_{\kappa}LM}({\rm LC:CG}),
\end{align}
where  
the operator ${\cal S}=\sum_{P}P$ ensures to extract a totally symmetric state.

\subsection{Search for linear-chain states}
\label{search.LC}

\begin{figure}
\begin{center}
  \epsfig{file=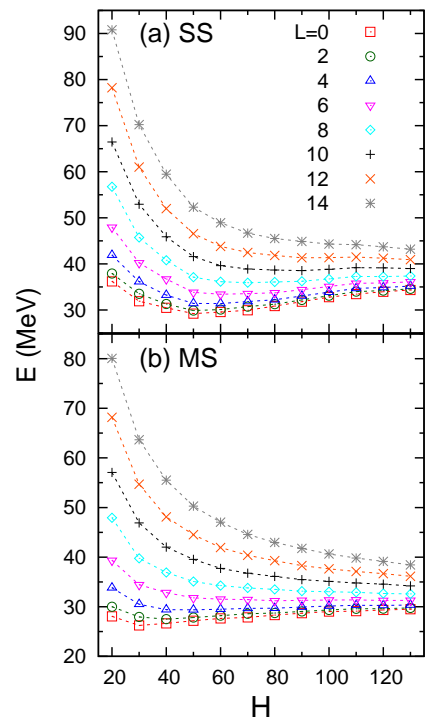,scale=1.1}
\caption{Energies of LC configurations with the angular momentum $L$ and 
positive parity. Panel (a) shows the lowest energy calculated 
in the SS model, whereas (b) the lowest energy in the MS model, respectively. Although 
it slightly depends on $L$, $r_{\rm rms}$ increases from about 
2.28 for $H=20$ to 5.62\,fm for $H=130$. }
\label{E-H.diag}
\end{center}
\end{figure}

First we compare the energies calculated with 
$ f^{\nu}_{S^H_{\kappa}LM}({\rm LC:CG})$ by varying $S_{\kappa}^H$ 
for a given $H$. This is called a single $S$ (SS) 
model. The lowest energy found in this model is plotted in panel (a) of 
Fig.~\ref{E-H.diag} as a function of $H$.  
The SS model finds a local energy minimum for $L=0, 2, 4, 6, 8$, 
and 10, although the minima for $L=8, 10$ are very shallow. The minima 
of $L=0,2,4$  appear at $H\approx 50$, which corresponds to the 
point-$\alpha$ rms radius, $r_{\rm rms}\approx 3.52$\,fm. The minimum of $L=6$ 
curve shifts to $H\approx 60$. For $L > 12$, no energy 
minimum appears in the region of $H< 130$, and the energy simply decreases with 
increasing $H$. 

As discussed above, the energy of the SS model with $L=0$ becomes 
a minimum at $H\approx 50$, in which four-$\alpha$ 
particles are 
positioned at $S_1=-4.295$, $S_2=-1.921$, $S_3=1.302$, $S_4=4.914$\,fm. 
This LC configuration is approximated by the CG that has $K=43$ and 
$a=0.874$. Before the 
boson symmetry is imposed, the relative distances calculated with this CG 
are $D_{12}=2.68$, $D_{13}=5.70$, $D_{14}=9.22$, $D_{23}=3.44$, $D_{24}=6.90$, 
$D_{34}=3.81$\,fm, respectively, which are all in very good agreement 
with those calculated from the LC state~(\ref{chain.wf}) 
\begin{align}
&\langle \Phi^{\nu}_{SLM}({\rm LC})|({\bm r}_i-{\bm r}_j)^2|\Phi^{\nu}_{SLM}({\rm LC})\rangle\notag \\
&=\frac{3}{\nu}(1-\delta_{i,j})
+\frac{1}{2}(S_i-S_j)^2\Big[1+\frac{L}{H}+\frac{i_{L+1}(H)}{i_L(H)}\Big].
\end{align}

Next we allow a mixing of various configurations, 
$\sum_{\kappa}C_{S_{\kappa}^H} f^{\nu}_{S^H_{\kappa}LM}({\rm LC:CG})$, while still keeping $H$  
fixed. This calculation named an MS model 
allows us to evaluate the extent to which the energy gain over the 
SS model is obtained by including the vibrational mode of the LC state. 
The lowest 
energy found in the MS model is plotted in panel (b) of Fig.~\ref{E-H.diag}.  
The MS model still presents a local energy 
minimum at $H\approx 30 \sim 40$ for $L=0, 2$, and at $H\approx 50$ for $L=4$.  
Although the minimum is found at $H\approx 80$ for $L=6$, the energy change 
is very little around that $H$ value. The energy curve of $L=8$ becomes 
flat with increasing $H$, and no minima appear for higher $L$ values. 
This suggests 
that the LC state with large $L$ value is probably not stable against 
the vibrational degree of freedom even under the LC restriction, 
but the minimum configuration 
found in the SS model tends to shift to larger rms size or to break into 
$\alpha$-particles. 

\begin{figure*}
\begin{center}
\epsfig{file=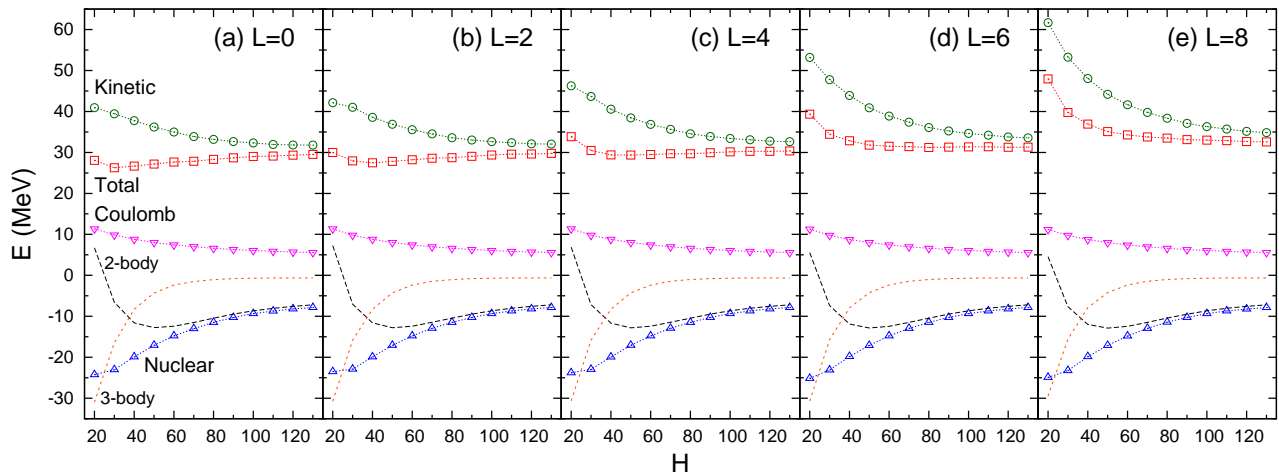,scale=0.96}
\caption{$H$-dependence of the contributions of the kinetic energy, the 
nuclear potential, and the Coulomb potential to the lowest energy LC states 
obtained in the MS model. The nuclear contribution consists of 
2-body ($V_{\rm 2B}$) and 3-body ($V_{\rm 3B}$) potential energies. Panels (a), (b), (c), (d), and (e) show $L^{\pi}=0^+, 2^+, 4^+$, $6^+$, and $8^+$ cases, respectively.}
\label{E-H.content}
\end{center}
\end{figure*}

The contributions of the kinetic energy, the 
nuclear potential ($V_{\rm 2B}+V_{\rm 3B}$), and 
the Coulomb potential to the lowest energy displayed in 
Fig.~\ref{E-H.diag}(b) are plotted in Fig.~\ref{E-H.content} for 
$L^{\pi}=0^+, 2^+, 4^+, 6^+, 8^+$
as a function of $H$. The kinetic energy 
gives a dominant contribution to the total energy. The nuclear and 
Coulomb potential contributions change very little with increasing $L$,  
whereas the kinetic energy contribution considerably depends on $L$. 
With increasing $L$, the kinetic-energy contribution rapidly increases 
as $H$ decreases, as expected, and cancels out the nuclear 
attractive contribution, leaving no energy minimum. It should be 
noted that the $V_{\rm 2B}$ and $V_{\rm 3B}$ potentials give 
an opposite contribution at very small $H$ while the $V_{\rm 2B}$ term 
plays a dominant role in the case of $H > 60$.  

The CG has the advantage that it can be applied to negative-parity states 
without any modification. To examine the possibility of finding a 
negative-parity LC state we study 
the $E-H$ diagram for $L^{\pi}=1^-$ in the same way as the positive-parity 
case. It turns out that the kinetic energy gets larger and the nuclear 
potential energy becomes much less attractive, which leads us to the conclusion 
that no negative-parity $4\alpha$ LC states exist.

Finally we mix various LC configurations with different $H$ values, 
$\sum_{H \kappa}C_{S^H_{\kappa}} f^{\nu}_{S^H_{\kappa}LM}({\rm LC:CG})$. 
This calculation is called an  
MH model. The configurations in the range of $H=20, 30,\ldots,130$ 
are included. The dimension of the Hamiltonian matrix is about 430. 
The energy gain obtained with the MH model, compared to 
the SS calculation, is very large, amounting to about 10 MeV. 
Table~\ref{MHresults} lists the result of the lowest LC states obtained 
in the MH model. 
The excitation energies, $E_x=E+E_{\rm th}$,  
of the $0^+, 2^+$, $4^+$, and $6^+$ LC states are predicted to 
be 32.9, 33.7, 35.3, and 37.2\,MeV, respectively. They follow 
the $(\hbar^2/2{\cal I})L(L+1)$-rule with 
$\hbar^2/2{\cal I}\approx 0.10$ or 0.12\,MeV if the $6^+$ state is excluded. 
This parameter is close to that of Ref.~\cite{yao14}, but considerably larger 
than those (0.06-0.08\, MeV) estimated 
in Refs.~\cite{chevallier67,ichikawa11,suhara14}. The energies of the 
second lowest LC states are 20.97, 21.72, 23.16, and 25.37\,MeV for 
$L^{\pi}=0^+, 2^+$, $4^+$, and $6^+$, respectively.

\begin{table}
\caption{Energies, given in units of MeV, of the lowest LC states 
from $4\alpha$ threshold predicted 
in the MH model and the contributions from the kinetic (K.E.), two-body potential 
($V_{2{\rm B}}$), three-body potential ($V_{3{\rm B}}$), and Coulomb potential 
($V_{\rm C}$) energies. The rms radius of $^{16}$O is estimated as 
$\sqrt{r_{\rm rms}^2+1.455^2}$ by taking into account the finite size of the 
$\alpha$-particle. } 
\label{MHresults}
\begin{center}
 \begin{tabular}{ccccccccccccccc}
\hline\hline
$L^{\pi}$ && $E$ && K.E. && $V_{2{\rm B}}$ && $V_{3{\rm B}}$ && $V_{\rm C}$ && $r_{\rm rms}$ (fm) && $\delta_2$ \\
\hline 
0$^+$ && 18.46 && 28.17 && $-$9.62 && $-$8.54 && 8.46 && 3.47 && 0\\
2$^+$ && 19.31 && 28.08 && $-$10.24 && $-$6.59 && 8.07 && 3.67 && $-$0.554\\
4$^+$ && 20.82 && 28.05 && $-$10.69 && $-$4.00 && 7.45 && 4.03 && $-$0.709\\
6$^+$ && 22.72 && 28.21 && $-$10.27 && $-$2.03 && 6.81 && 4.43 && $-$0.783\\
\hline\hline
\end{tabular}
\end{center}
\end{table}

The contributions of the kinetic energy and the potential energies to 
$E$ show an interesting contrast. Both contributions of 
the kinetic energy and the two-body nuclear potential are only weakly 
dependent on $L$. On the other hand, the contributions of 
the three-body nuclear and Coulomb potentials alter significantly 
as a function of $L$, 
that is, they follow the change of the rms radius that increases with $L$.
The $L(L+1)$ rotational spectrum 
of the LC states is therefore mainly due to the three-body nuclear and 
Coulomb potentials, which are both long-range pieces of the Hamiltonian. 
This is in sharp contrast to the rotation of a rigid-body 
where the kinetic energy should play a primary role in forming the $L(L+1)$ 
pattern. 

Table~\ref{MHresults} lists the $\delta_2$ value~(\ref{delta2}) as well. 
The limiting value of $\delta_2$ with the CG is 
$-2L/(2L+3)$, as shown in Appendix~\ref{me.quadrupole}.  The MH model gives $\delta_2$ 
close to that limit for each $L^+$ state, indicating a large quadrupole 
deformation. 
Assuming $\langle x^2\rangle=\langle y^2 \rangle$ 
valid for an axial symmetric shape, the calculated $\delta_2$ value   
suggests the ratio of the major radius to the minor radius  
of the LC state as
\begin{align}
\frac{\sqrt{\langle y^2 \rangle}}{\sqrt{\langle z^2 \rangle }}
=\sqrt{\frac{2-\delta_2}{2(1+\delta_2)}}\approx 1.69,\ 2.16,\ 2.53
\end{align}
for $L=2,\ 4, \ 6$, respectively. Note that the ratio approaches 
$\sqrt{L+1}$ in the limit of $\delta_2\to -2L/(2L+3)$. If we assume the 
obtained LC state to have such intrinsic density that is 
constant inside an axially symmetric spheroid with the 
quadrupole deformation parameter $\beta$, we may estimate $\beta$ 
from the following equation
\begin{align}
\frac{\sqrt{\langle y^2 \rangle}}{\sqrt{\langle z^2 \rangle }}
=\frac{1+\sqrt{\frac{5}{16\pi}}2\beta}{1-\sqrt{\frac{5}{16\pi}}\beta}.
\end{align} 
The resulting values of $\beta$ are respectively 0.59, 0.88, 1.07 for 
$L=2,\ 4, \ 6$, indicating very large deformation. 

The CG is based on a spherical representation and includes no explicit 
deformation parameters. Nevertheless the resulting wave functions 
are found to represent very large deformation. This is primarily 
made possible by the use of very large $K$ values. A usual approach 
is to explicitly include some parameters relevant to the deformation. 
For instance, it is shown in Ref.~\cite{suhara14} that the LC state has 
very large overlap with the rotating state projected from the 
intrinsically deformed configuration
\begin{align}
e^{-\frac{\nu}{2}R_x^2-\frac{\nu'}{2}R_y^2-\frac{\nu}{2}R_z^2},
\label{deformed.hypradial}
\end{align}
where $R_x^2$ and $R_y^2$ are defined in exactly the same way as $R_z^2$, and  
$\nu'$ is taken much smaller than $\nu$, typically 0.027\,fm$^{-2}$, to 
embody the shape elongated along the $y$ direction. Note that 
Eq.~(\ref{deformed.hypradial}) is obtained by deforming the 
hyperradial Gaussian, $e^{-\frac{\nu}{2}R^2}=e^{-\frac{\nu}{2}\widetilde{\bm \rho}\Lambda \bm \rho}$.

\begin{figure}
\begin{center}
\epsfig{file=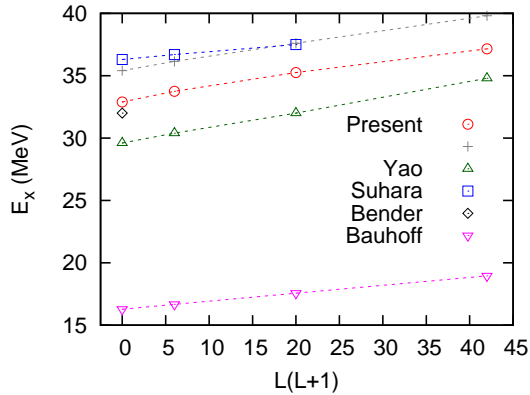,scale=1.2}
\caption{Comparison of the excitation energies of the 4$\alpha$ LC states 
as a function of $L(L+1)$. Open circles and crosses denote the lowest and 
the second lowest LC states obtained in the MH model. Data are taken from 
Yao~\cite{yao14}, Suhara~\cite{suhara14}, Bender~\cite{bender03}, and 
Bauhoff~\cite{bauhoff84}.}
\label{Ex.LC}
\end{center}
\end{figure}

The excitation energies of the LC states obtained by several models 
are compared in Fig.~\ref{Ex.LC}. Except for Brink's $\alpha$-cluster model 
result~\cite{bauhoff84}, the bandhead of the LC states is predicted to be 
much higher than that speculated in Ref.~\cite{chevallier67} and targeted 
experimentally~\cite{freer95,curtis16}. Our bandhead energy (the open circle) 
is close to that estimated by a generator 
coordinate approach using the Skyrme-Hartree-Fock (HF)+BCS model~\cite{bender03} and 
falls between those energies calculated by 
a covariant density functional calculation~\cite{yao14} and a generator 
coordinate treatment of the Brink's wave functions~\cite{suhara14}. Our 
result is very satisfactory in view of the phenomenological treatment of the 
Hamiltonian. 
In contrast to a cranked HF calculation~\cite{ichikawa11} that 
reports the stabilized LC states with $L=$13\,-18,
our calculation finds no 
local energy minima at the LC configurations rotating with such high 
angular momenta, consistently with Refs.~\cite{yao14,suhara14}. The second 
lowest LC states of our calculation are drawn by crosses in the figure. 
They are in fair agreement with the result of Ref.~\cite{suhara14}.

\section{Conclusions}
\label{conclusion}

The single-particle wave function that is angular-momentum projected 
from the Gaussian wave packet provides spatially localized configurations. 
We have shown that the linear-chain configuration projected 
from a product of the Gaussian wave packets 
can be to high accuracy approximated by the correlated Gaussians 
if their parameters $A, u$ and $K$ are chosen under the condition developed 
here.  Although the correlated Gaussians 
have been known for the last twenty years, a particular choice of $K$ 
has led to those configurations which have strongly deformed intrinsic 
shape but nevertheless are eigenfunctions of the angular momentum. 
 
The present formulation makes it possible to perform 
calculations of variation-after-projection type. Combined with the advantage 
that the needed matrix elements can be analytically obtained with the 
correlated 
Gaussians, we have studied a system of four $\alpha$-particles in order 
to examine 
possible existence of the linear-chain states in $^{16}$O. The 
two-$\alpha$ and three-$\alpha$ phenomenological potentials are 
set to reproduce the ground state and the first excited 2$^+$ state of $^{12}$C 
reasonably well. The energies of the chain configurations are calculated 
as a function of the angular momentum as well as the size of the 
four-$\alpha$ system. The full calculation taking into account 
the vibration of the $\alpha$-particles and the 
extension of the chain length finds the possibility of 
the chain states with $0^+, 2^+$, and $4^+$. The case with $6^+$ may be 
marginal, and no 
possibility of chain states is observed for $J\geq 8$. The bandhead energy 
of the chain states is about 33\,MeV from the ground state of $^{16}$O. 
Though those chain states are found to follow a rotational sequence of 
$J(J+1)$, its physical aspect is significantly different from the case of 
an ideal rigid-body rotation. The combined 
effect of the long-range pieces of the Hamiltonian is mainly responsible 
for its pattern.

One of the important issues is the stability 
of the linear chain state. To discuss the stability, one has to 
take account of the coupling of the linear chain configurations with other more general 
configurations. We have to prepare 
various configurations and to perform calculations that include them together 
with the linear chain states. Those configurations can be expressed by 
the correlated Gaussians that are specified by flexible parameters, $A, u, K$. A work along 
this direction is interesting and will be reported elsewhere.

\acknowledgments

One of the authors (Y.S.) would like to thank N. Itagaki for 
useful discussions that were made possible by a  
visiting program (January 2017) of Yukawa Institute for Theoretical Physics, Kyoto University. He also heartily thanks L. Tomio for the generous invitation 
(September to December 2015) to Universidade Federal do ABC, 
where an early stage of the work was done. 
This work was in part supported by JSPS KAKENHI Grant No. 15K05072.

\appendix

\section{Glossary of acronyms}
\label{glossary.acronym}

\begin{center}
\begin{tabular}{lllll}
\hline\hline
Acronym && Full word && First occurrence \\
\hline 
CG && correlated Gaussian && Eq.~(\ref{def.CG}) \\
c.m. && center-of-mass && Sec.~\ref{gwprepre} \\
GWP && Gaussian wave packet && Eq.~(\ref{gwp}) \\
HO && harmonic-oscillator && Sec.~\ref{amp.gwp} \\
LC && linear-chain && Sec.~\ref{intro} \\
LPG && locally peaked Gaussian && Eq.~(\ref{nodelessho}) \\
MH && mixed $H$ && Sec.~\ref{search.LC} \\
MS && mixed $S$  && Sec.~\ref{search.LC} \\
rms && root-mean-square && Sec.~\ref{arrangements} \\
SG && shifted-Gaussian && Eq.~(\ref{radial.sgf}) \\
sp && single-particle && Sec.~\ref{intro} \\
SS && single $S$  && Sec.~\ref{search.LC} \\
VAP && variation after projection && Sec.~\ref{shape} \\
VBP && variation before projection && Sec.~\ref{shape} \\
\hline\hline
\end{tabular}
\end{center}

\section{Glossary of symbols}
\label{glossary.symbol}

\begin{center}
\begin{tabular}{lll}
\hline\hline
Symbol && Definition \\
\hline 
\vspace{1mm}
$G_{\kappa}$ && $\Big(\frac{\Gamma(\kappa +\frac{3}{2})}{2\sqrt{\pi}}\Big)^{\frac{1}{2}}$ \\
\vspace{1mm}
$\gamma_{KK'L}(x)$ && $\sum_{n=0}^{{\rm min}(K,K')}\frac{K!\,K'!\,\Gamma(L+\frac{3}{2})}{n!\,(K-n)!\,(K'-n)!\,\Gamma(n+L+\frac{3}{2})}x^n$\\
\vspace{1mm}
${\cal N}^{uA}_{KL}$ && $\frac{1}{(2K+2L+1)!!}\sqrt{\frac{4\pi (2L+1)!!}{\gamma_{KKL}(1)}}
\Big(\frac{{\rm det}A}{\pi^{N-1}}\Big)^{\frac{3}{4}}  $ \\
   && $ \times \Big(\frac{1}{2}\widetilde{u}A^{-1}u\Big)^{-\frac{2K+L}{2}}$  \\
\vspace{1mm}
$G_{KL}(y,z)$ && $\frac{{\pi}^{\frac{1}{4}}}{\sqrt{2\gamma_{KKL}(1)}}
 \sum_{n=0}^K\frac{ K!\, \sqrt{\Gamma(L+\frac{3}{2})}\, y^{K-n}
\,(\frac{z}{2})^{2n+L}} {n!\,(K-n)!\,\Gamma(n+L+\frac{3}{2})}$ \\
\vspace{1mm}
$b_{l}(x)$ && $\big( i_l(x)\, e^{-x}\big)^{\frac{1}{2}}$\\
\vspace{1mm}
$O_l(\nu s, ak)$ && $\frac{1}{2}\Big(\frac{\eta}{2}\Big)^{\frac{l}{2}} 
\Big(\frac{2a}{\nu +a}\Big)^k \Big(\frac{2\sqrt{\nu a}} {\nu +a}\Big)^{l+\frac{3}{2}} 
\frac{\Gamma(k+l+\frac{3}{2})}{b_{l}(\eta)\Gamma(l+\frac{3}{2}) G_{2k+l}}$ \\
  && where $\eta=\frac{1}{2}\nu s^2$ \\
\vspace{1mm}
$B_{KL}$ && $\frac{4\pi (2K+L)!}{(2K)!!\, (2K+2L+1)!!}$ \\
\vspace{1mm}
$g(\alpha,\bm e;u,A,\bm \rho)$ && $\exp\big(-\frac{1}{2}\widetilde{\bm \rho}A\bm \rho+\alpha \bm e\cdot(\widetilde{u}\bm \rho)\big)$ \\
\hline\hline
\end{tabular}
\end{center}
Here $\Gamma$ is the gamma function. 

\section{Function $zi_{l+1}(z)/i_l(z)$}
\label{func.fzl}

\begin{figure}
\begin{center}
\epsfig{file=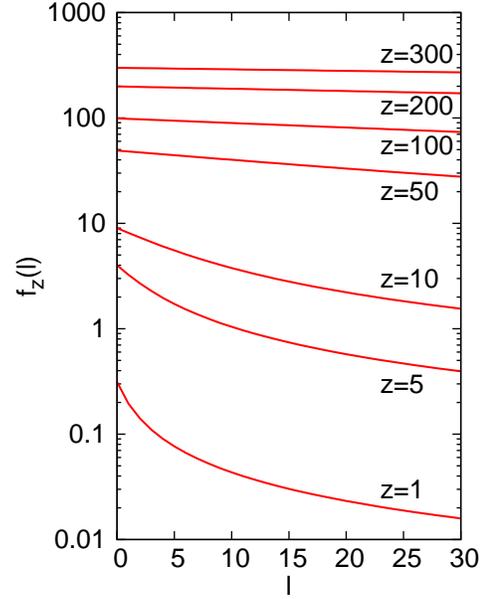,scale=1.0}
\caption{Plots of $f_z(l)$ as a function of $l$ for some values of $z$.  }
\label{mspbessel}
\end{center}
\end{figure}

As seen in Eqs.~(\ref{me.r2.kine.SG}) and (\ref{radius.kinetic.LC}), the 
function $f_z(l)=zi_{l+1}(z)/i_l(z)$ appears in the matrix elements involving 
the angular-momentum projected GWP.  
Figure~\ref{mspbessel} plots $f_z(l)$ for some values of $z$. 
We find numerically that the following inequality holds:
\begin{align}
f_z(l) \geq \sqrt{\big(l+\frac{3}{2}\big)^2+z^2}-l-\frac{3}{2}.
\label{approx.f}
\end{align}
Actually the right-hand side of the inequality is a good approximation 
to  $f_z(l)$. The relative error becomes largest at $l=0$ for any $z$, 
and its maximum relative error becomes largest, about 
8\%, at $z\approx 3$. Note that 
$f_z(l) \approx z^2/(2l+3)$ for $z \ll l$, while $f_z(l) \approx 
z-l-\frac{3}{2}$ for $z \gg l$. 

If a rigid-body approximation works well, we expect that the rms radius 
is constant 
and the kinetic energy is proportional to $l(l+1)$. This is translated to the 
following behavior of $f_z(l)$: $f_z(l)+l+z$ is $l$-independent, 
whereas $f_z(l)+l-z$ is proportional to $l(l+1)$. 
Numerical check indicates that this expectation very much depends on $z$. 
For $z=20$, the rigid-body approximation is reasonable  
up to $l=20$. With increasing $z$ the approximation 
works better up to larger $l$. E.g., with $z=40$, the 
approximation works well up to about $l=40$, and with  
$z=60$ it significantly improves. 

\section{Matrix element of $\widetilde{\bm \rho}\Omega {\bm \rho}$} 
\label{me.rhoOmegarho}
 
The operator $\widetilde{\bm \rho}\Omega {\bm \rho}$ is scalar and quadratic in $\bm \rho$, where the matrix $\Omega$ may not be necessarily symmetric.
We obtain its matrix element between the CGs following the procedure 
of Ref.~\cite{suzuki98}: 
\begin{align}
&\langle f^{uA}_{KLM}|\widetilde{\bm \rho}\Omega {\bm \rho}|f^{vB}_{K'LM}\rangle
=\frac{{\cal N}^{uA}_{KL}{\cal N}^{vB}_{K'L}}{B_{KL}B_{K'L}}\notag \\
&\quad \times \int \!\! \int d{\bm e} \, d{\bm e}'
Y_{LM}^*(\hat{\bm e})Y_{LM}(\hat{\bm e'})\Big(
\frac{d^{2K+L+2K'+L}}{d\alpha^{2K+L}d\alpha'^{2K'+L}}\notag \\
&\quad \times \langle g(\alpha, \bm e;u,A,\bm \rho)|\widetilde{\bm \rho}\Omega {\bm \rho}|g(\alpha',\bm e';v,B,\bm \rho)\rangle \Big)_{\alpha=\alpha'=0}.
\label{cg.me.xOx}
\end{align}
The matrix element between the generating functions reads 
\begin{align}
&\langle g(\alpha, \bm e;u,A,\bm \rho)|\widetilde{\bm \rho}\Omega {\bm \rho}|g(\alpha',\bm e';v,B,\bm \rho)\rangle\notag \\
&=\Big(\frac{\pi^{N-1}}{\det C}\Big)^{\frac{3}{2}}
e^{p\alpha^2+p'\alpha'^2+q\alpha \alpha'\bm e\cdot \bm e'}\notag \\
&\times \Big[3R(\Omega)+P(\Omega)\alpha^2+
P'(\Omega)\alpha'^2+Q(\Omega)\alpha \alpha'\bm e\cdot \bm e'\Big],
\label{xomegax.m.e}
\end{align}
where $C=\frac{1}{2}(A+B)$ and the various coefficients are  
\begin{align}
&p=\frac{1}{4}\widetilde{u}C^{-1}u,\ \ \  p'=\frac{1}{4}\widetilde{v}C^{-1}v,\notag \\
&q=\frac{1}{4}(\widetilde{u}C^{-1}v+\widetilde{v}C^{-1}u),\notag \\
&P(\Omega)=\frac{1}{4}\widetilde{u}C^{-1}\Omega C^{-1}u,\ \ \ P'(\Omega)=\frac{1}{4}\widetilde{v}C^{-1}\Omega C^{-1}v,\notag \\
&Q(\Omega)=\frac{1}{4}(\widetilde{u}C^{-1}\Omega C^{-1}v+\widetilde{v}C^{-1}\Omega C^{-1}u),\notag \\
&R(\Omega)=\frac{1}{2}{\rm Tr} C^{-1}\Omega.
\end{align}
Performing the differentiation and integration  
in Eq.~(\ref{cg.me.xOx}) leads to the following result:
\begin{align}
&\langle f^{uA}_{KLM}|\widetilde{\bm \rho}\Omega {\bm \rho}|f^{vB}_{K'LM}\rangle \notag \\
&=\frac{1}{\sqrt{\gamma_{KKL}(1)\gamma_{K'K'L}(1)}}{\bar{t}}^{\frac{L}{2}} \notag \\
&\times \Big(\frac{\sqrt{\det AB}}{\det C}\Big)^{\frac{3}{2}}\Big(\frac{\widetilde{u}C^{-1}u}{\widetilde{u}A^{-1}u}\Big)^K
\Big(\frac{\widetilde{v}C^{-1}v}{\widetilde{v}B^{-1}v}\Big)^{K'}\notag \\
&\times \sum_{n=0}^{\min (K,K')}\frac{K!\,K'!\, \Gamma(L+\frac{3}{2})}{n!\, (K-n)!\, (K'-n)!\,\Gamma(n+L+\frac{3}{2})}t^n\notag \\
&\times \Big[3R(\Omega)+(K-n)\frac{P(\Omega)}{p}+(K'-n)\frac{P'(\Omega)}{p'}\notag \\
&\ \  +(L+2n)\frac{Q(\Omega)}{q}\Big],
\end{align}
where
\begin{align}
t=\frac{(\widetilde{u}C^{-1}v)^2}{(\widetilde{u}C^{-1}u)(\widetilde{v}C^{-1}v)}, \ \ \ 
{\bar{t}}=\frac{(\widetilde{u}C^{-1}v)^2}{(\widetilde{u}A^{-1}u)(\widetilde{v}B^{-1}v)}.
\end{align}

The diagonal matrix element takes a simple form:
\begin{align}
&\langle f^{uA}_{KLM}|\widetilde{\bm \rho}\Omega {\bm \rho}|f^{uA}_{KLM}\rangle \notag \\
&=\frac{3}{2}{\rm Tr} A^{-1}\Omega+(2K+L)\frac{\widetilde{u}A^{-1}\Omega A^{-1}u}{\widetilde{u}A^{-1}u}.
\end{align}
The matrix element of $r_{ij}^2$ is obtained by putting 
$\Omega=\omega^{(ij)}\widetilde{\omega^{(ij)}}$. 
The formula~(\ref{rms.kinetic.cg}) for $R^2$ is 
obtained by replacing $\Omega$ with $\Lambda$ in the above equation.

In some cases one may want to calculate 
the mean deviation of $r_{ij}^2$ from its mean value, {\it i.e.}, 
$\langle (r_{ij}^2-D_{ij}^2)^2\rangle=\langle r_{ij}^4\rangle -D_{ij}^4$. 
It is calculated by using the approximation
\begin{align}
\langle f^{uA}_{KLM}|e^{-\epsilon\widetilde{\bm \rho}\Omega {\bm \rho}}|f^{uA}_{KLM}\rangle 
\approx 1-\epsilon D_{ij}^2+\frac{\epsilon^2}{2} \langle r_{ij}^4\rangle+\ldots.
\end{align}
The left-hand side of this equation is nothing but the expectation value 
of the Gaussian-type potential. By taking small $\epsilon$ value, $\langle r_{ij}^4\rangle$ is easily obtained.

\section{Matrix element of quadrupole moments} 
\label{me.quadrupole}

We consider the matrix element
\begin{align}
\langle f^{uA}_{KLM}|R_z^2 |f^{vB}_{K'LM}\rangle
=\langle f^{uA}_{KLM}|\widetilde{\zeta}\Lambda \zeta |f^{vB}_{K'LM} \rangle,
\end{align}
where $\zeta=(\zeta_i)$ is the $(N-1)$-dimensional column vector comprising 
the $z$ component of ${\bm \rho}_i$. Although we need the case with $M=L$, we here 
work out for a general case. 
The basic step for calculating the matrix element is the same as that of 
$\widetilde{\bm \rho}\Lambda \bm \rho$. The matrix element between the generating 
functions is easily obtained because they are factorized in $x, y, z$ components: 
\begin{align}
&\langle g(\alpha, \bm e; u, A,\bm \rho)|\widetilde{\zeta}\Lambda \zeta |g(\alpha', 
{\bm e}'; v,B,\bm \rho \rangle\notag \\
&=\Big(\frac{\pi^{N-1}}{\det C}\Big)^{\frac{3}{2}}
e^{p\alpha^2+p'\alpha'^2+q\alpha \alpha'\bm e\cdot \bm e'}\notag \\
&\times  \Big[R(\Lambda)+P(\Lambda)\alpha^2e_z^2
+P'(\Lambda)\alpha'^2e_z'^2 +Q(\Lambda)\alpha \alpha'e_ze_z'\Big],
\label{zetalambda}
\end{align}
where $e_z$ is the $z$ component of the unit vector $\bm e$, $e_z=\sqrt{4\pi/3}Y_{10}(\hat{\bm e})$, and likewise $e_z'=\sqrt{4\pi/3}Y_{10}(\hat{\bm e'})$. 
The matrix element~(\ref{zetalambda}) takes a form similar 
to that of Eq.~(\ref{xomegax.m.e}), but because of 
the difference in the tensor character of $\widetilde{\zeta}\Lambda \zeta$, 
it contains explicit dependence on $e_z$ and $e_{z}'$ as well. 
The manipulation needed to obtain the desired matrix 
element thus involves a slightly lengthy procedure compared to the case of 
$\widetilde{\bm \rho}\Lambda \bm \rho$, leading to  
the following result:
\begin{align}
&\langle f^{uA}_{KLM}|\widetilde{\zeta}\Lambda {\zeta}|f^{vB}_{K'LM}\rangle \notag \\
&=\frac{1}{\sqrt{\gamma_{KKL}(1)\gamma_{K'K'L}(1)}}{\bar{t}}^{\frac{L}{2}} \notag \\
&\times \Big(\frac{\sqrt{\det AB}}{\det C}\Big)^{\frac{3}{2}}\Big(\frac{\widetilde{u}C^{-1}u}{\widetilde{u}A^{-1}u}\Big)^K
\Big(\frac{\widetilde{v}C^{-1}v}{\widetilde{v}B^{-1}v}\Big)^{K'}\notag \\
&\times \sum_{n=0}^{\min (K,K')}\frac{K!\,K'!\, \Gamma(L+\frac{3}{2})}{n!\, (K-n)!\, (K'-n)!\,\Gamma(n+L+\frac{3}{2})}t^n\notag \\
&\times \Big[R(\Lambda)+(K-n)\frac{P(\Lambda)}{p}c_{LM}^{(0)}
+(K'-n)\frac{P'(\Lambda)}{p'}c_{LM}^{(0)} \notag \\
& \ \  +\frac{Q(\Lambda)}{q}c_{nLM}^{(1)}\Big],
\end{align}
where
\begin{align}
&c_{LM}^{(0)}=\frac{1}{2L+3}+\frac{2(L^2-M^2)}{(2L-1)(2L+3)},\notag \\
&c_{nLM}^{(1)}=2nc_{LM}^{(0)}+\frac{L^2-M^2}{2L-1}.
\end{align}

The diagonal matrix element with $M=L$ reads 
\begin{align}
&\langle f^{uA}_{KLL}|\widetilde{\zeta}\Lambda {\zeta}|f^{uA}_{KLL}\rangle \notag \\
&=\frac{1}{2}{\rm Tr}A^{-1}\Lambda+\frac{2K}{2L+3}
\frac{\widetilde{u}A^{-1}\Lambda A^{-1}u}{\widetilde{u}A^{-1}u},
\end{align}
and therefore $\delta_2$ defined by 
Eq.~(\ref{delta2}) is found to be 
\begin{align}
&\delta_2=-\frac{2L}{2L+3}\frac{2K+L+\frac{3}{2}}
{2K+L+\frac{3}{2}\kappa}
\end{align}
with
\begin{align}
\kappa=\frac{\widetilde{u}A^{-1}u}{\widetilde{u}A^{-1}\Lambda A^{-1}u}{\rm Tr} A^{-1}\Lambda.
\end{align}
For the choice of $A=a\nu \Lambda$, Eq.~(\ref{A.choice}), used to approximate 
the LC configuration with the CG, $\kappa$ reduces to $N-1$ independent of 
$u$. Then $\delta_2$ approaches $-2L/(2L+3)$ in the limit of 
$K\to \infty$.

\end{document}